\documentclass{aa}
\usepackage[varg]{txfonts}

\usepackage{graphicx}
\usepackage{natbib}
\usepackage{multirow}
\usepackage{amsmath}
\usepackage{amssymb}
\usepackage{url}

\newcommand{\hd}[0]{HD~189733}
\newcommand{\cahk}[0]{\ion{Ca}{II} \mbox{H \& K}}
\newcommand{\catr}[0]{\ion{Ca}{II}}
\newcommand{\nad}[0]{\ion{Na}{I}}
\newcommand{\ha}[0]{\ion{H}{$\alpha$}}
\newcommand{\hb}[0]{\ion{H}{$\beta$}}
\newcommand{\hg}[0]{\ion{H}{$\gamma$}}
\newcommand{\hde}[0]{\ion{H}{$\delta$}}
\newcommand{\he}[0]{\ion{H}{$\epsilon$}}
\newcommand{\hei}[0]{\ion{He}{I} \mbox{D3}}

\newcommand{\mg}[0]{\ion{Mg}{I}}
\newcommand{\hel}[0]{\ion{He}{I}}
\newcommand{\fe}[0]{\ion{Fe}{II}}
\newcommand{\ergs}[0]{\mbox{erg\,s$^{-1}$}}

\begin{document}

\title{Time-resolved UVES observations of a stellar flare on the planet 
host \hd\ during primary transit\thanks{Based on observations 
made with ESO Telescopes at the La Silla Paranal Observatory under programme ID 089.D-0701(A)}}

\author{T. Klocov\'a\inst{1,2} \and S. Czesla\inst{1} \and S. 
Khalafinejad\inst{1} \and U. Wolter\inst{1} 
   \and J.\,H.\,M.\,M. Schmitt\inst{1}}
\institute{Hamburger Sternwarte, Universit\"at Hamburg, Gojenbergsweg 112, 21029
Hamburg, Germany
\and
Astronomical Institute ASCR, Fri\v{c}ova 298, 25165 Ond\v{r}ejov, Czech 
Republic\\ \email{klocova@asu.cas.cz}}

\date{Received ... / Accepted ... }

\abstract{HD 189733 is an exoplanetary system consisting of a transiting 
hot Jupiter and an active K2V-type main sequence star. Rich manifestations 
of a stellar activity, like photometric spots or chromospheric flares were repeatedly observed in this system in optical, UV and X-rays.}
{We aim to use VLT/UVES high resolution ($R=60\,000$) echelle spectra to study a stellar flare.
}{We have performed 
simultaneous analyses of the temporal evolution in several chromospheric stellar lines, namely, the
\cahk\ lines (3933, 3968\,\AA), \ha\ (6563\,\AA), \hb\ (4861\,\AA), \hg\ (4341\,\AA), \hde\ (4102\,\AA), \he\ 
(3970\,\AA), the \catr\ infrared 
triplet line (8498, 8542 and 8662\,\AA), and \hei\ (5875.6\,\AA). 
Observations were carried out with a time resolution of approximately\,1\,min 
for a duration of four hours, including a complete
planetary transit.}
{We determine the energy released during the flare in all
studied chromospheric lines combined to be 
about $8.7\times 10^{31}$\,erg, which puts this event at the upper end of flare energies 
observed on the Sun.
Our analysis does not reveal any
significant delay of the flare peak observed in the Balmer and \cahk\ lines, 
although we find a clear difference in the temporal evolution of these lines.
The \hei\ shows additional absorption possibly related to the flare event.
Based 
on the flux released in \cahk\ lines during the flare, we estimate the
soft X-ray flux emission to be $7\times 10^{30}$\,erg.}{ 
The observed flare can be 
ranked as a moderate flare on a K-type star and confirms a 
rather high activity level of 
\hd\ host star. The cores of the studied chromospheric lines 
demonstrate the same behavior and let us study the flare 
evolution. We demonstrate that the activity of an exoplanet 
host star can play an important role in the detection of 
exoplanet atmospheres, since these are frequently discovered 
as an additional absorption in the line cores. A possible 
star-planet interaction responsible for a flare occurrence 
during a 
transit can neither be confirmed nor ruled out.}


\titlerunning{Flare on \hd}
\authorrunning{T. Klocov\'a et al.} 
\maketitle

\section{Introduction}\label{intro}

Because of its proximity to Earth (d~=~19.3\,pc) and apparent 
brightness ($V=7.7$\,mag), \hd\
has become one of the most studied exoplanetary systems
\citep[e.g.,][and references therein]{Madhusudhan2014}. Its host star,
\hd\,A, is a cool star of spectral type K2V, which shows the usual
signatures of magnetic activity. 
Photometric observation made with HST by \citet{Pont2007} show flux residuals 
during a transit, which can be assigned to a spot 
complex on the stellar surface (with the size in longitude of at least $80\,000$\,km)
that is occulted by the transiting planet. 
An extensive spectroscopic study of \hd\ by 
\citet{Boisse2009} shows activity in several of the ``classic'' activity 
indicators such as \cahk, \ha, and \hei. 
With their time series spanning 
more than 300\,days \citet{Boisse2009} also 
detect a modulation in the chromospheric and photospheric (spotted surface) activity tracers with a 
12~day period, identified as the rotation period of the star; this
period had previously been estimated by \citet{Henry2008} as
$P_{\mathrm{rot}} = 11.953 \pm 0.009$\,days,  
based on an extensive long-term observing campaign covering 18 complete cycles over approximately 400\,days.

The first discovery of transits of a hot Jupiter planet orbiting 
around \hd\,A is reported
by \citet{Bouchy2005}, and a little later \citet{Bakos2006} report the discovery of
an additional M-type stellar companion \hd\,B in the system in a 216\,AU orbit. 
The basic characteristics of the star and the planet are summarized in 
Table~\ref{tab:Properties}. Currently \hd\,A is one of the most active exoplanetary host stars 
known. \citet{Gray2003} measure a strength of the
chromospheric activity  (described in more detail in Sect.~\ref{Sec:rhk}) 
of $R'_{\mathrm{HK}}=-4.553$ and \citet{Knutson2010} find a 
similar value of $R'_{\mathrm{HK}}=-4.501$. These values classify \hd\ as an 
active star, according to the classification suggested by \citet{Henry1996}.

X-ray emission from \hd\,A has been detected with the EXOSAT and ROSAT satellites
(at a time when \hd\,A was not known as a planet host), 
and \citet{cutispoto2000} quote typical X-ray luminosity levels
in the range 1.5-2.5 $\times$ 10$^{28}$ erg/s from these observations.  
Comparing this value to the X-ray luminosity distribution function observed for G and K dwarfs
in the solar neighborhood \citep{schmitt1995, schmitt1997} 
shows that -- not surprisingly -- \hd\,A is located in the high-luminosity part 
of the
X-ray luminosity distribution function and thus \hd\,A is a very active star 
also from the X-ray point of view.

The existence of a massive exoplanet on a close-in orbit around an
active star makes \hd\, a very promising target for the study of 
star-planet interactions. The possible influence on the chromospheric and coronal 
activity caused by the presence of a close-in Jupiter-like planet is 
still a matter of ongoing studies and debates.  For example,
\citet{Pillitteri2010} present a study of an X-ray flare event observed 
with {\it XMM-Newton}, occurring 3\,ks after the end of the planetary secondary 
transit (at orbital phase $\sim 0.54$); their analysis suggests flare temperatures
of $\sim 10$\,MK, similar to temperatures of solar flares, yet the size of the 
flaring loop is estimated to be on the order of the stellar radius.
Using again XMM-Newton, the same group of authors report a further flare observed near secondary
transit (at phase 0.52; \citet{Pillitteri2011}) and \citet{Pillitteri2014} 
report the occurrence of a third flare at phase 0.65, with similar characteristics 
as the previous two flares, with a loop of length of four stellar radii at the location of the flare.

Using HST/STIS, \citet{Lecavelier2012} find significant variations in an 
evaporating atmosphere of \hd\ imprinted on the Ly $\alpha$ line of \hd\, and 
suggest that these changes could have been caused by an increased flow of 
charged particles released during an X-ray flare detected by the
Swift satellite 9\,hours before the HST observation.  Recently
\citet{Pillitteri2015} presented new FUV observation with HST/COS of \hd , observing
\hd\, during five consecutive HST orbits, covering the orbital phase of 0.5-0.63 and 
detecting signatures of enhanced activity in the form of variability in
the lines from the elements Si, C, N, and O.

In this paper we present ground-based observations of \hd\, using VLT/UVES and 
covering a primary transit. Directly in the middle of the transit,
a flare occurred, that was followed with a high temporal cadence thus allowing
a detailed analysis of this particular stellar flare. 
Since the flare itself 
causes changes several times bigger in the line cores  
than the exoplanetary transit, in this paper we focus exclusively on 
the flare analysis.
Flares are thought to be caused by a sudden energy release in the stellar corona  
and observed as a brightening across the whole electromagnetic spectrum 
\citep[for review see 
e.g.][]{Benz2008}. Their occurrence  is unpredictable in nature and their observation 
requires either long term monitoring or some kind of luck (or bad luck, 
especially if one intends to use the data for a different purpose).
The energy released in solar flares is generally between $10^{29} - 
3\times 10^{32}$\,ergs, but 
also more energetic events seem to have been observed in 
several cases \citep[reported e.g. by][]{Notsu2013}.  The energies of
stellar flares can exceed these solar values by orders of magnitude.
The observed stellar chromospheric activity related to a flare arises 
essentially from the transfer of energy from the corona to the chromosphere, which
occurs via accelerated particles (electrons) traveling 
along magnetic field lines. The subsequent heating of the chromosphere leads to an 
increasing level of excitation of \ha\ and other 
chromospheric lines. 

In our analysis, we focus on the 
temporal evolution of 
chromospheric lines visible in the optical and the determination of energy released via these lines. 
In Sect.~\ref{Sec:obsandred} we describe the 
observations and data reduction. In Sect.~\ref{results} we describe the 
analysis of the data, the behavior of the 
chromospheric lines, and their energy budget. In Sect.~\ref{Sec:discussion} we discuss our results and 
in Sect.~\ref{Sec:conclusions} 
we give the conclusions.

\begin{table}
  \caption{Properties of the host star \hd\,A - top part, and of transiting planet \hd \,b - bottom part of the table.
  \label{tab:Properties}}
  \begin{tabular}[h]{l l l}
  \hline\hline
  Property & Value & Source$^a$ \\ \hline
  $T_{\mathrm{eff}}$ [K] & $5040\pm 50$ & T \\
  $\log$(g[cm\,s$^{-2}$]) & $4.587^{+0.014}_{-0.015}$ & T \\
  $[$Fe/H$]$ & $-0.04\pm 0.08$ & B/T \\
  $v\sin(i)$ [km\,s$^{-1}$] & $3.5\pm 1$ & B \\
  $M_{\mathrm{s}}$ [M$_{\odot}$] & $0.806\pm0.048$ & T \\
  $R_{\mathrm{s}}$ [R$_{\odot}$] & $0.756\pm0.018$ & T \\
  Age [Gyr] & 0.6 & M \\
  Distance [pc] & $19.5\pm 0.3$ & H \\
  \hline
    $P_{\mathrm{orb}}$\,[days] & $2.21857567 \pm 1.5\times10^{-7}$ & A\\
  $M_{\mathrm{p}}$ [M$_{\mathrm{J}}$]& $1.144^{+0.057}_{-0.056}$ & T\\
  $R_{\mathrm{p}}$ [R$_{\mathrm{J}}$]& $1.138\pm 0.027$ & T\\
  $T_0$ $\mbox{MJD}_{\mathrm{HJD,UTC}}$ & $56109.26164$ & this work\\
  $a$ [AU]& $0.03099^{+0.0006}_{-0.00063} $ & T\\
  $i$ [deg] & $85.58\pm 0.06 $& T \\ \hline
  \end{tabular}
  \tablefoot{$^a$\; B:\citet{Bouchy2005}, T:\citet{Torres2008},
  A:\citet{Agol2010}, M:\citet{Melo2006}, H:Hipparcos \citet{vanLeeuwen2007}.}
\end{table}

\section{Observations and data reduction}
\label{Sec:obsandred}
Here, we briefly summarize the most relevant aspects of the data
processing; a more detailed description of the observations and data 
reduction can be found in \citet{Czesla2015}.
The spectroscopic echelle observations of \hd\ were obtained during the night of
July 1/2, 2012, with the UVES spectrograph mounted at the VLT-UT2 telescope in
Chile. The data set consists of $244$ individual spectra, covering four hours of
observation.  The exposure time for the first 29 spectra was 30\,s and 
the rest was exposed for 45\,s, resulting in a temporal resolution of 
approximately 1\,min. The observations cover an entire primary 
transit of \hd\,b ($\sim 110$\,min) plus some pre- and post-transit time. For
the data reduction, the UVES pipeline in version $5.2.0$ was used. We applied
the standard reduction procedure, and also flux
calibrated the spectra using the master response files.

\begin{figure}[h]
  \includegraphics[width=0.5\textwidth]{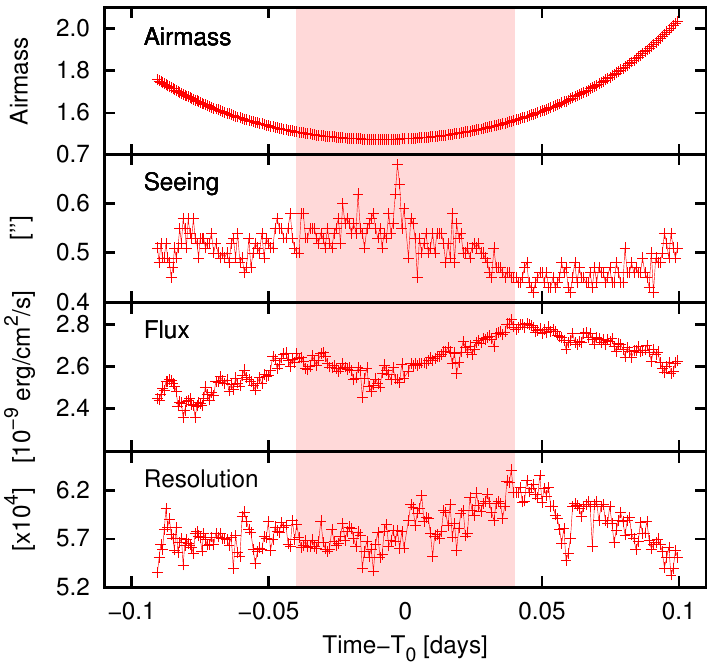}
  \caption{Observing conditions and evolution of spectral resolution. 
  {\it From top to bottom:} Temporal evolution of the airmass, seeing, spectral
  flux measured on a blue chip, and instrumental resolution. 
Red crosses represent single data points.
  The shaded area indicates the transit duration and the transit mid-time,
  $T_0$, refers to $\mbox{MJD}_{\mathrm{HJD,UTC}} = 56109.26164$. This value is 
used in all the following figures.
  \label{fig:SeeingAirmass}}
\end{figure}

Figure~\ref{fig:SeeingAirmass} shows the temporal evolution of airmass, seeing,
the observed flux on the blue chip, and the spectral resolution; the
duration of the transit is also indicated. 
During the first half of the observations,
the seeing varies between $0.5$'' and $0.6$'' and decreases
afterwards. The slit width was 1'' for the blue and 0.7'' for the red arm, still
larger but on the same order as the seeing disk.
We studied the temporal changes in the line profiles using telluric spectral
lines. In particular, we selected 15 water lines in the range from $6288$ to
$7308$\,\AA\ and derive the median instrumental resolution by measuring their
width. The spectral resolution varied during the observation as a result of
the changing slit illumination when the seeing disk moves in the slit
\citep[see Fig.\,\ref{fig:SeeingAirmass} and][]{Czesla2015}.

\citet{Czesla2015}
compare the integrated flux from the individual chips
with model spectra \citep{Castelli2004, Kurucz1970} and a blackbody
spectrum and find that they generally agree well (see their Table~4);
they further attribute the 
remaining variations in the observed flux of about $20$\% (also shown in
Fig.~\ref{fig:SeeingAirmass}) to slit losses and the residual influence of
airmass and atmospheric conditions. Clearly, the residual flux evolution
defines a lower error limit for any flux measurement based on these spectra.

Figure~\ref{fig:SeeingAirmass} shows a visible evolution in the width of the
telluric lines during the observing night, related to changing
observing conditions. Since variability in the telluric lines can affect the 
results of a spectroscopic line analysis, and, in particular, also affects
the regions around the activity sensitive \ha\ and \catr\ IRT triplet lines,
a correction of the science spectra for telluric lines is necessary.
To remove these lines we used the \texttt{Molecfit} 
code\footnote{\url{http://www.eso.org/sci/software/pipelines/skytools/molecfit}}
\citep{Kausch2015, Smette2015}. \texttt{Molecfit} works by fitting
synthetic telluric transmission spectra to user-specified sections of the
astronomical data. Based on the thus derived atmospheric conditions, a synthetic
transmission spectrum is then calculated for the entire spectral range. As fitting
ranges we selected several telluric molecular bands occurring in the observed
spectrum and fit for the three molecules H$_2$O, O$_2$ and O$_3$, since these
are the most relevant telluric contributors in the optical range. The resulting
model transmission spectrum is then used to correct the pipeline-reduced
spectrum.  The residuals in the spectrum after applying 
the molecfit correction are 
within 5\% for the strongest water lines.

\section{Analysis of chromospheric lines}\label{results}

The cores of strong spectral lines are often formed in the chromosphere and
can thus be used to study the processes occurring in this atmospheric layer.
Our UVES spectra cover a wide wavelength range (3760--4980\,\AA, 5700--7530\,\AA,
 and 7650--9460\,\AA) with a spectral resolution of 60\,000.
Therefore,
the chromosphere of \hd\ can be studied in many chromospheric lines, 
formed at different atmospheric heights. In our analysis, we focused on the
\cahk\ lines (3933, 3968\,\AA), 
the \catr\ IR triplet lines (8498, 8542, 8662\,\AA), the Balmer series with \ha\
(6563\,\AA), \hb\ (4861\,\AA), \hg\ (4341\,\AA), \hde\ (4102\,\AA), and \he\
(3970\,\AA), and the \hei\ line (5876\,\AA).

\subsection{Chromospheric activity indicators}

\noindent
The \cahk\label{intro:cahk} resonance lines are among the most
pronounced spectral features in spectra of late type stars. They
result from transitions from $4\,
^{2}\textrm{S}_{1/2}$ to $4\, ^{2}\textrm{P}_{1/2}^{\textrm{o}}$ and 
$4\,^{2}\textrm{S}_{1/2}$ to $4\, ^{2}\textrm{P}^{\textrm{o}}_{3/2}$,
respectively and
have been regularly used in chromospheric studies, such as the
long-term, ground-based survey of stellar activity initiated by O.C.\,Wilson
to analyze the temporal evolution of activity levels and detect
analogs of the Sun's 11~year activity cycle in other stars
\citep[e.g.,][]{Wilson1968, Baliunas1995}.
Several indices such as the Mount-Wilson S-index have been introduced to
quantitatively describe the \cahk\ line emission. In our analysis, we focus 
on the $\log R'_{\mathrm{HK}}$ index \citep{Noyes1984}, described in more detail
in Sect.\,\ref{Sec:rhk}.

The same ion gives rise to
the \catr\ IR triplet lines, which are strong spectral features in the
near infrared region in the spectra of late type stars. They are
formed in the lower levels of the chromosphere and arise from the transitions
$4 ^{2}\textrm{P}^{\textrm{o}}_{3/2}$ to $3 ^{2}\textrm{D}_{3/2}$ at 8498\,\AA, 
$4 ^{2}\textrm{P}^{\textrm{o}}_{3/2}$ to $3 ^{2}\textrm{D}_{5/2}$ at 8542\,\AA,
and $4 ^{2}\textrm{P}^{\textrm{o}}_{1/2}$ to $3 ^{2}\textrm{D}_{3/2}$ at
8662\,\AA. The \catr\ IRT lines were recognized and used as activity indicators for example, by
\citet{Linsky1979} and \citet{Chmielewski2000}.

The cores of the Balmer series lines, most importantly, the {\it \ha} line, are
also commonly used activity indicators. The \ha\ line has regularly been used in
ground-based studies of solar flares \citep[e.g.,][]{Canfield1984, Leka1993,
lang2009} and flares observed on other stars
\citep[e.g.,][]{Fuhrmeister2008,Lalitha2013}. In the Sun 
it is formed in the 
middle of the chromosphere \citep{Vernazza1981}.

The \hei\ line, which is actually a multiplet resulting from the
transitions between the $2 ^{3}\textrm{P}^{\textrm{o}}$ and the $3
^{3}\textrm{D}$ terms, is extremely weak in the quiescent solar photosphere,
where it cannot be formed because of its high excitation potential of about
$21$~eV \citep[e.g.,][]{Landman1981}.
On the Sun, the \hei\ line appears in absorption 
in eruptive filaments, flare ejecta, and weak flares and turns into 
emission only in intense flares \citep[][Sect.~11.3]{Feldman1983, Zirin1988}.
It shows strong spatial correlation with solar plage regions
\citep{Landman1981}. While the details of its formation are still debated
\citep[e.g.,][]{Athay1960, Zirin1975, Andretta1995, Lanzafame1995}, its absence
in the quiet photospheres of the Sun and other late-type stars makes it an
excellent indicator of magnetic activity.

The relevant He levels can be populated by photoionzation
followed by recombination (P-R mechanism) or collisional excitation in sufficiently hot and
dense plasma \citep[e.g.,][]{Andretta1995, Lanzafame1995}.
To be observable in emission in solar flares, the line requires a temperature of
$>15\,000$\,K and a density in excess of $10^{14}$\,cm$^{-3}$ \citep{Feldman1983}.

\subsection{Temporal evolution of the chromospheric activity index $\log R'_{\mathrm{HK}}$}
\label{Sec:rhk}
To study the temporal evolution and energetics of the \cahk\ lines, we
determine the chromospheric index $\log R'_{\mathrm{HK}}$ for the entire
sequence of observations.
The $R'_{\mathrm{HK}}$ index is
defined as the ratio of the chromospheric surface flux
in the \cahk\ line cores, $F'_{\mathrm{HK}}$, and the bolometric
flux of the star through
\begin{equation}
  R'_{\mathrm{HK}} = \frac{F'_{\mathrm{HK}}}{\sigma T_{\mathrm{eff}}^{4}} \, ,
  \label{eq:rhkprime}
\end{equation}
where $\sigma$ is the Stefan-Boltzmann constant and $T_{\mathrm{eff}}$ the
effective temperature.
The $\log R'_{\mathrm{HK}}$ index is based on the Mount Wilson
S-index ($S_{\mathrm{HK}}$), from which it mainly differs by excluding photospheric emission and
by its normalization \citep[for $S_{\mathrm{HK}}$, see e.g.][]{Middelkoop1982,
Wright2004}.

\begin{figure}[h]
\begin{center}
  \includegraphics[width=0.5\textwidth]{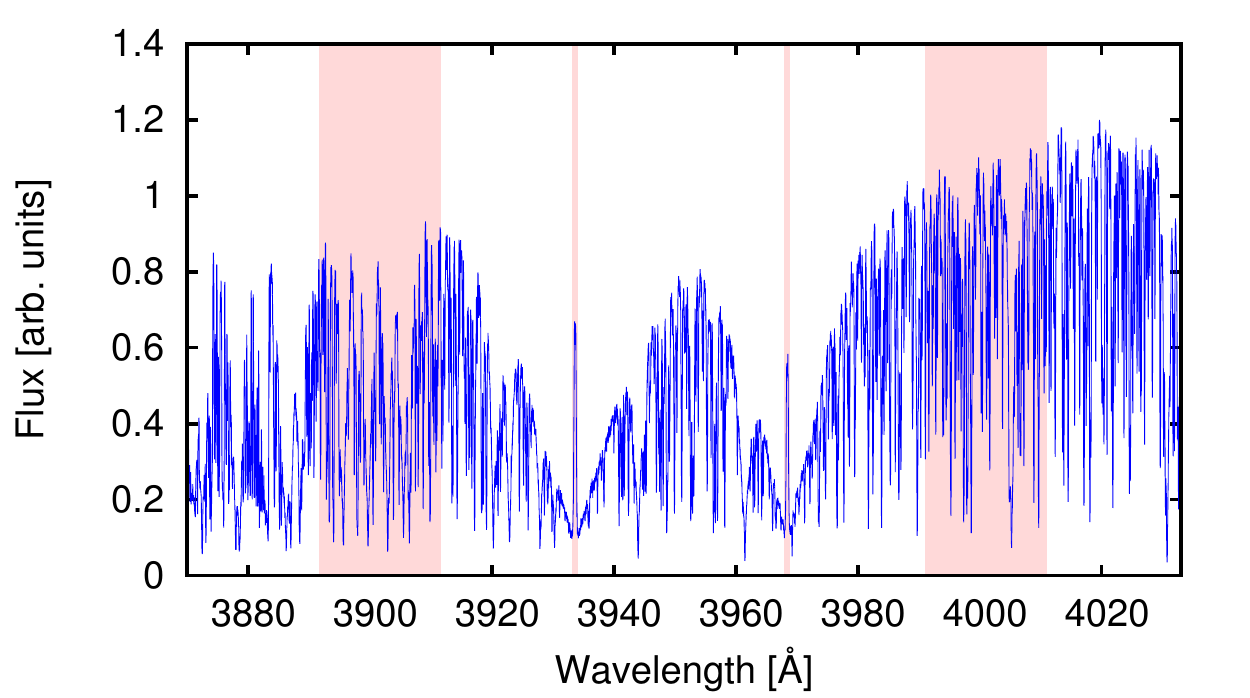}
  \caption{Spectrum around the \cahk\ lines. Shaded areas represent the two
  reference bands and the two 1\,\AA\ wide intervals centered on the \cahk\ line
  cores used for the determination of the $\log R'_{\mathrm{HK}}$ index.
  \label{fig:hkpassband}}
\end{center}
\end{figure}

To derive the $\log R'_{\mathrm{HK}}$ index, we first
determined $S_{\mathrm{HK}}$
following the
calibration procedure described by \citet{Melo2006}. In particular, we
measured the fluxes in two 1\,\AA\ wide
intervals centered on the H and K line cores, $F_{\mathrm{H}}$ and
$F_{\mathrm{K}}$, and determined
the fluxes in two reference bands $C_{\mathrm{H}}$ ($3891.67-3911.67$~\AA) and
$C_{\mathrm{K}}$ ($3991.067-4011.067$~\AA). The selected passbands are depicted in Fig.\,\ref{fig:hkpassband}.
Subsequently, we calculated their so-called $S_{\mathrm{US}}$ index
defined as ($F_{\mathrm{H}}$ + $F_{\mathrm{K}}$)/($C_{\mathrm{H}}$ +
$C_{\mathrm{K}}$)
and converted it into the Mount Wilson S-index based on the relation
\citep{Melo2006}:
\begin{equation}
  S_{\mathrm{US}}=0.06111\times S_{\mathrm{HK}} - 0.00341.
\end{equation}
Following this, we transformed $S_{\mathrm{HK}}$ into
$\log R'_{\mathrm{HK}}$ following the procedure described
by \citet{Noyes1984} and \citet{Wright2004}, applying 
a value of $B-V = 0.93$ taken from \citet{Koen2010}.

\begin{figure}
\begin{center}
\includegraphics[width=0.5\textwidth]{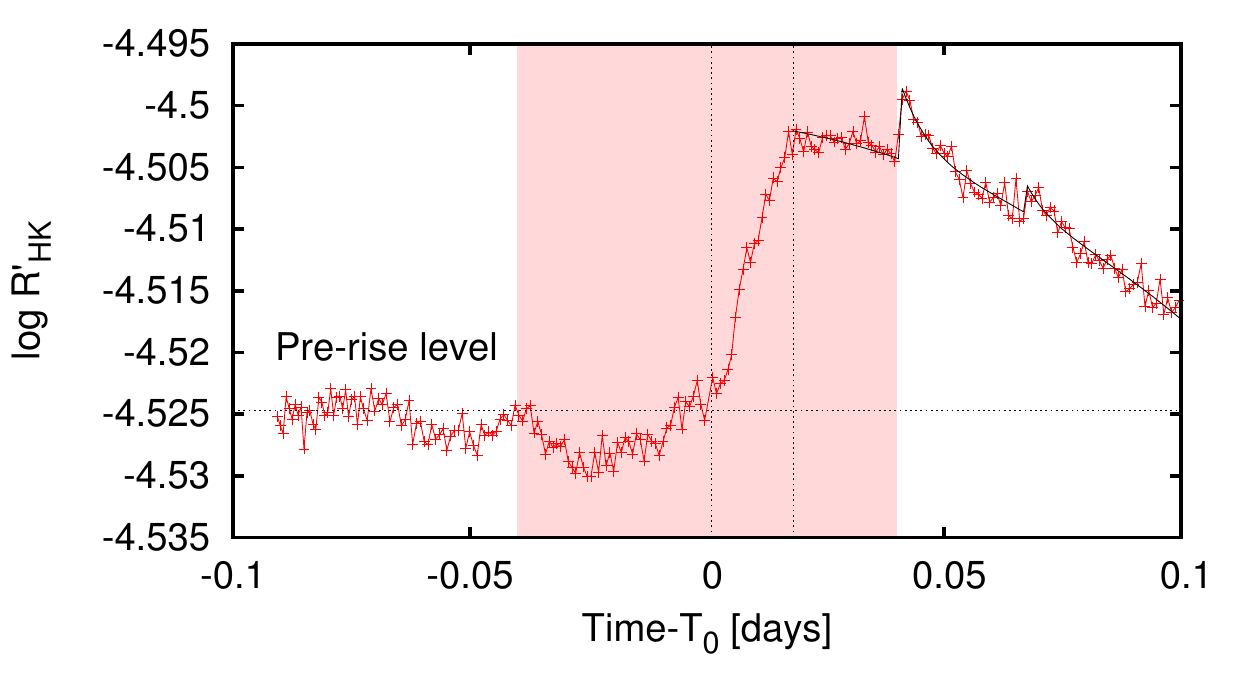}
\caption{Temporal evolution of the chromospheric activity index $\log
R'_{\mathrm{HK}}$. The primary rise and peak phases (dashed, vertical lines),
the adopted pre-flare level (horizontal dashed line), and the best fit of the
gradual decay phase along with the secondary and tertiary flares are indicated.
The center of the $x$ axis (time=0\,days) corresponds 
to the middle of 
the transit ($T_{0}$\,$[\mathrm{HJD,UTC}] = 2456109.26164$). The shaded pink area 
indicates the duration of an exoplanetary transit, the 1\,$\sigma$ error for the 
data points varies between 3-4\%.
\label{fig:rhkIndex}}
\end{center}
\end{figure}

Figure~\ref{fig:rhkIndex} shows the
temporal evolution of the $\log R'_{\mathrm{HK}}$ index for our UVES observations,
which indicates a relatively quiescent phase during the first half of the observations, where only
little variability is seen, followed by an abrupt rise in the middle of the 
transit, which subsequently
turns into a slow decay, continuing beyond the end of our
observing run. We dub the rise and subsequent decay the ``primary event''.
The decay phase of the primary event is interrupted by a smaller, secondary and
a tertiary event with a shorter rise and decay phase seen $\sim56$~min
and $\sim95$~min after the start of a primary event, respectively.
All three events are characterized by a short rise and a comparably long decay
phase, which is typical for stellar and solar flares \citep[e.g.][]{Benz2008}. 
Therefore, we interpret the observed temporal behavior in terms of three stellar
flares. For all three, Fig.\,\ref{fig:rhkIndex} allows us to 
distinguish three flare phases, namely the rise (or impulsive) phase,
the peak of the flare, and a gradual decay phase.

A quantitative analysis of the flares requires defining a quiescent
level. Given the observed variability pattern, a truly quiescent phase is hard
to identify. After visual inspection of Fig.\,\ref{fig:rhkIndex}, we used
the beginning of the observations as the quiescent phase. In
particular, we classified the mean of the first 40 spectra as our 
quiescent reference spectrum. The corresponding quiescent $\log
R'_{\mathrm{HK}}$ value of $-4.525$ is represented by a horizontal line
in Fig.~\ref{fig:rhkIndex}.

Similarly, the onset of the primary event cannot be
uniquely identified.
In Fig.~\ref{fig:rhkIndex}, we indicate the definition adopted for the
beginning of the flare rise phase and the instant of peak flux. These times 
are determined as an intersection of two linear fits to visually selected 
sections of the light curve neighboring the flare peak and start.
According to our
definition, the flare starts $2$\,min after mid-transit time and shows a  
$24$\,min long rise phase.

Along with the data, Fig.~\ref{fig:rhkIndex} shows a fit to the decay part of
the light curve, which consists of a third-order polynomial to represent the
decay phase of the primary flare. The
secondary and tertiary events can be fit using an instantaneous rise phase
followed by an exponential decay on top of the polynomial background.
Although the decay of the primary flare is not well described by an
exponential, it can be characterized
by an e-folding time.
This quantity is defined as the time after which the flare flux
drops to $37$\% compared to its maximum. We estimate an
e-folding time of approximately $111$~min for the primary event.
According to our modeling, the secondary event
reaches a peak level of $-4.499$ in $R'_{\mathrm{HK}}$ and shows a decay time of $8.7$~min.
The tertiary flare shows a slightly shorter e-folding time of $6.5$~min.
All the derived values are summarized in Table~\ref{tab:FlareRhk}.

Figure~\ref{fig:rhkIndex} also shows a dip 
in the light curve visible at the 
beginning of the transit and interrupted by the stellar 
flare in the middle of the transit. This dip can probably 
be attributed to the center-to-limb variation, which is 
described in more detail in \citet{Czesla2015}.

\begin{figure}[h]
\begin{center}
\includegraphics[width=0.5\textwidth]{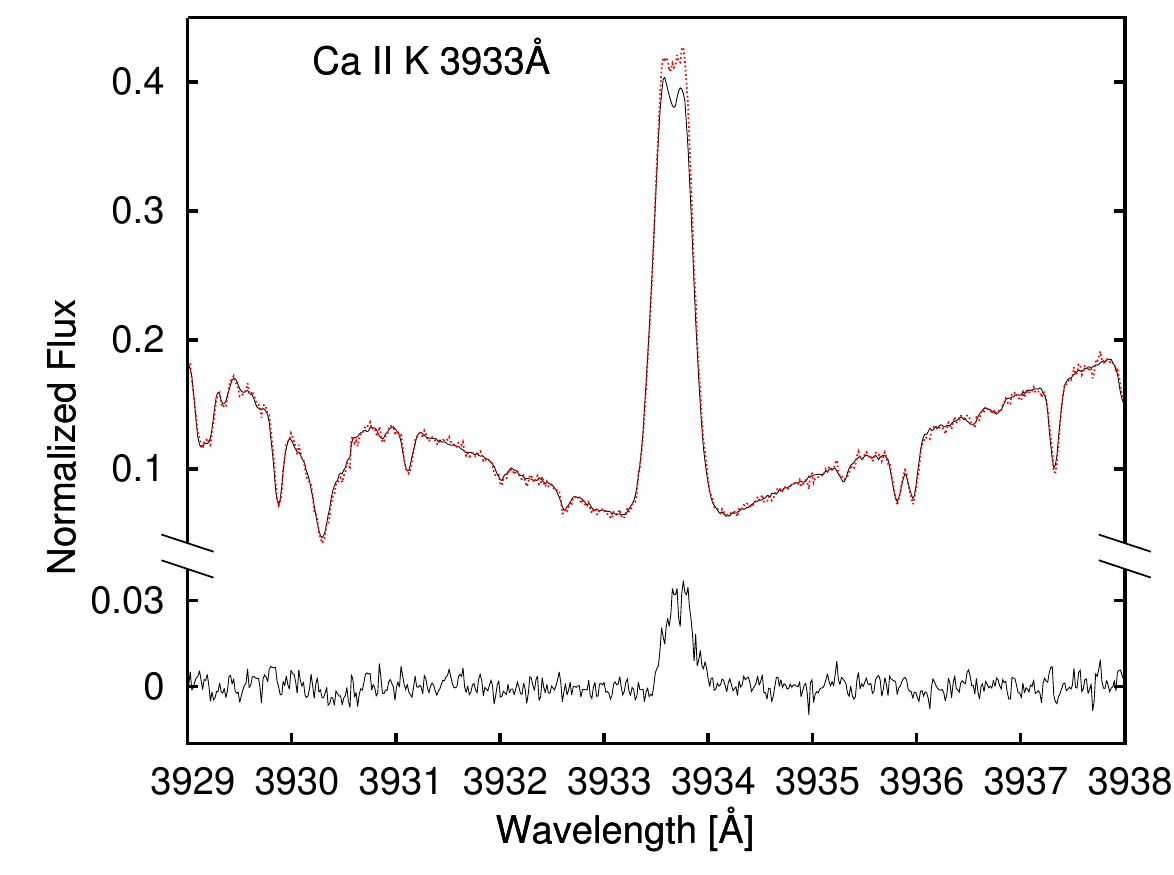}
\caption{Activity in the core of the \catr\ K line (3933\,\AA). {\it Upper
part:} Quiescent reference spectrum (solid black line) and spectrum
observed during the flare peak (dotted red line).
{\it Bottom curve:} Flux residuals after subtracting the quiescent spectrum
from the flare spectrum (solid black line).
\label{fig:excesCaii}}
\end{center}
\end{figure}

The rise in $\log R'_{\mathrm{HK}}$ associated with the
primary flare can directly be observed in the spectrum. In
Fig.\,\ref{fig:excesCaii}, we show the quiescent reference spectrum of the \catr~K line along with a
spectrum from the most active flare phase. Clearly, the latter exhibits stronger
\cahk\ line cores.

\begin{table}
\begin{center}
\caption{Characteristics of the flares determined from the $R'_{\mathrm{HK}}$
index.}\label{tab:FlareRhk}
\begin{tabular}{ll}
\hline
\hline
Parameter & Value\\
\hline
$S_{\mathrm{PF}}$ [MJD-56109] & $0.263 \pm 0.002$\\
$\Delta t_{\mathrm{rise}}$ [min]& 24.2\\
$M_{\mathrm{PF}}$ [MJD-56109] & $0.2798 \pm 0.0008$\\[3pt]
$M_{\mathrm{SF}}$ [MJD-56109] & $0.3052^{+0.0048}_{-0.0043}$\\[3pt]
$M_{\mathrm{TF}}$ [MJD-56109] & $0.3193^{+0.0090}_{-0.0093}$\\[3pt]
 $E_{\mathrm{Tot}}$[$10^{31}$\,erg] & 1.26 \\
$\tau_{\mathrm{PF}}$ [min]& $111.14$\\
$\tau_{\mathrm{SF}}$ [min]& $11.0^{+1.8}_{-1.9}$\\[3pt]
$\tau_{\mathrm{TF}}$ [min]& $7.6^{+1.9}_{-1.6}$\\[1pt]
$L_{\mathrm{MP}}$ [$10^{27}$\,erg s$^{-1}$]&  2.06\\
$L_{\mathrm{MS}}$ [$10^{26}$\,erg s$^{-1}$] & 5.175 \\
$L_{\mathrm{MT}}$ [$10^{25}$\,erg s$^{-1}$] &  3.055 \\
\hline
\end{tabular}
\tablefoot{$S_{\mathrm{PF}}$ - start of the primary 
flare with $1\sigma$ 
confidence limits; $\Delta t_{\mathrm{rise}}$ - duration of the primary rise phase; 
$M_{\mathrm{PF}}$, $M_{\mathrm{SF}}$, $M_{\mathrm{TF}}$ - instant of peak flux of the primary, secondary,
and tertiary flare;
$E_{\mathrm{Tot}}$ - energy released during a whole event; 
$\tau_{\mathrm{PF}}$, $\tau_{\mathrm{SF}}$, $\tau_{\mathrm{TF}}$ - e-folding
times of the primary, secondary, and tertiary flare;
$L_{\mathrm{MP}}$, $L_{\mathrm{MS}}$, $L_{\mathrm{MT}}$ - peak luminosity
of the primary, secondary, and tertiary flare in \cahk. Calculation of $1\sigma$ 
errors for $S_{\mathrm{PF}}$ and $M_{\mathrm{PF}}$ are 
based on \citet{Filliben1972}. Errors of $M_{\mathrm{SF}}$,$M_{\mathrm{TF}}$, 
$\tau_{\mathrm{SF}}$ and $\tau_{\mathrm{TF}}$ are determined by MCMC method.
}
\end{center}
\end{table}

\subsection{Time resolved analysis of chromospheric flux residuals}
\label{Sec:exces_cores}
In this section we present an analysis of the chromospheric core emission
observed in the \catr\ IRT lines, the lines of the Balmer series, and the \hei\
line, along with an alternative analysis of the emission in the \cahk\ lines
based directly on the flux calibrated spectra and not relying on the conversion of
chromospheric indices.
To study the effect of the flare on chromospherically sensitive lines, we
derive the temporal evolution of the flux residuals in their line cores.
As detailed in Sect.~\ref{Sec:rhk}, we use the mean of
the first 40 spectra as a quiescent reference spectrum to which the remaining
spectra are compared to detect residual emission similar to the core flux
residuals already indicated in Fig.\,\ref{fig:excesCaii}.

\begin{figure*}[t]
\begin{center}
  \includegraphics[width=\textwidth]{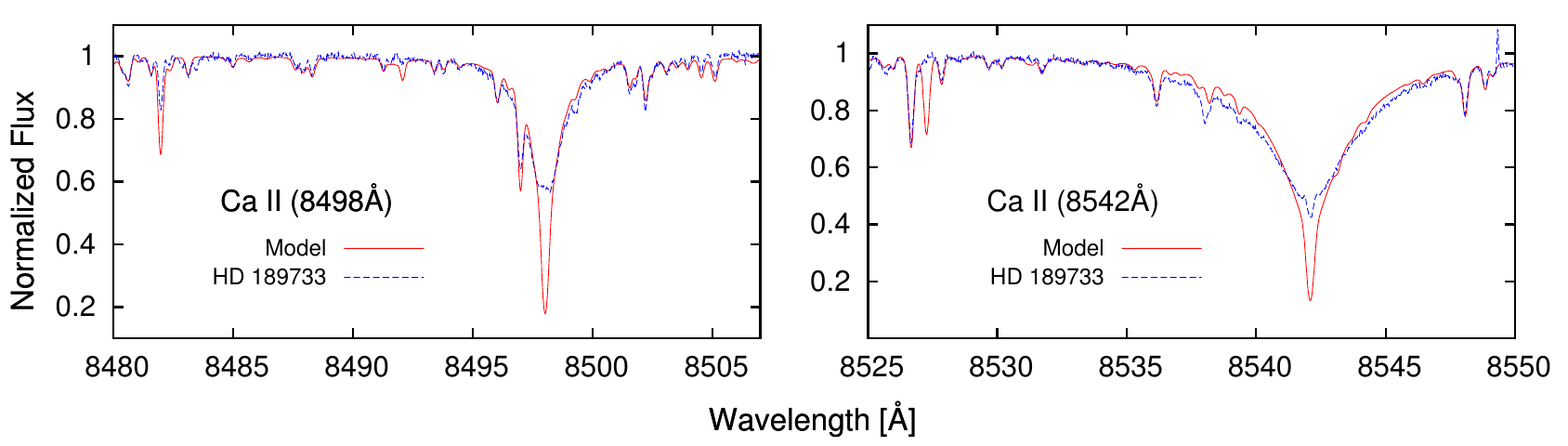}
  \caption{Comparison of the observed first (8498\,\AA) and second (8542\,\AA) Ca II IRT line profiles (blue dashed line) with a
photospheric model (red solid line). The model was created using the SPECTRUM
program \citep{Gray1994} based on a Kurucz stellar atmosphere model for
$T_{\mathrm{eff}}=5000$\,K, $\log g = 4.5$, and solar metallicity
\citep{Castelli2004}.
  \label{fig:irt_sun}}
\end{center}
\end{figure*}

A comparison of the \catr\ IRT lines in our quiescent reference spectrum with
synthetic model spectra shown in Fig.\ref{fig:irt_sun} clearly demonstrates that
the central parts of the lines show a noticeable fill-in also in the
quiescent phase. This is comparable to but less 
pronounced than the cores of
the \cahk\ lines and a consequence of the generally high level of stellar activity in \hd. 

\begin{figure}
\begin{center}
\includegraphics[width=0.5\textwidth]{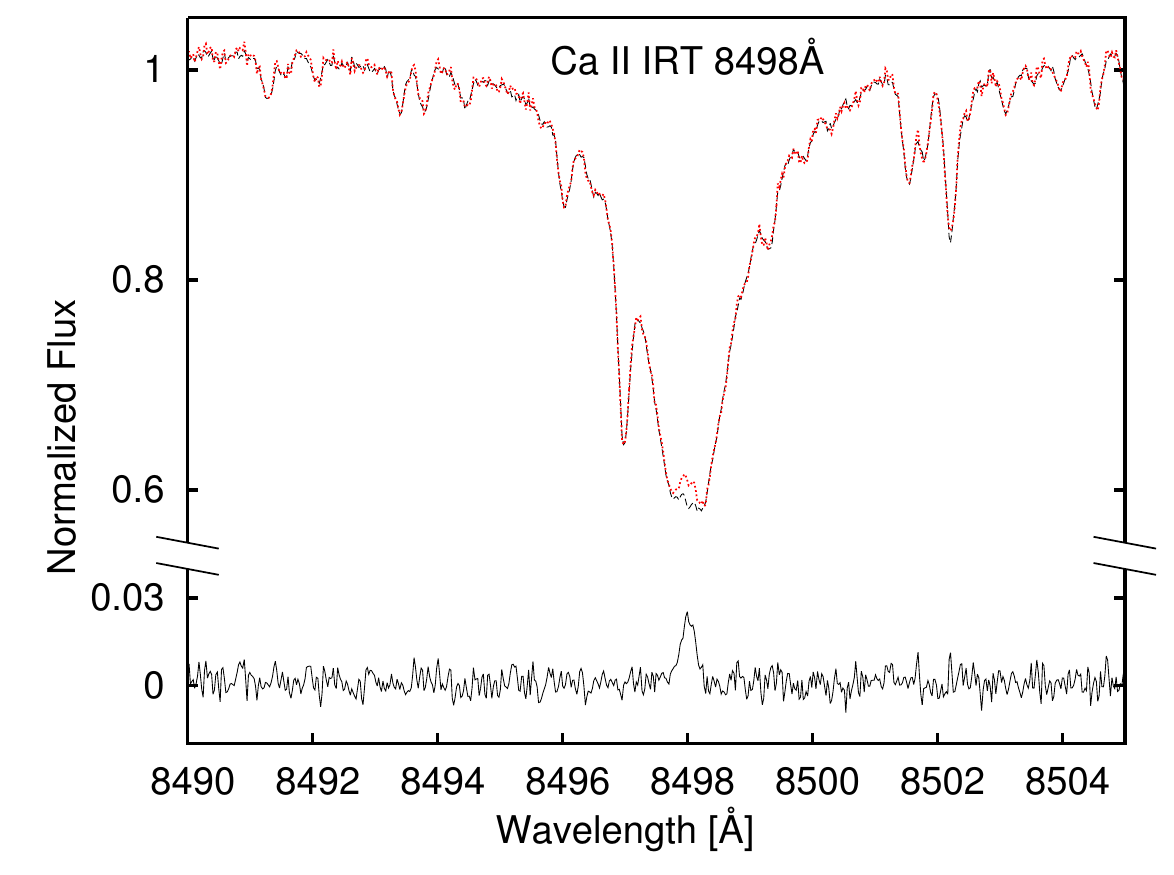}
\caption{Activity in the core of the first \catr\ IRT line (8498\,\AA). {\it Upper part:} dotted red line - spectrum observed during the 
activity peak (second half of the observation); dashed black line - quiescent spectrum determined as an average value of 40 spectra 
obtained at the beginning of the observation. {\it Bottom curve:} solid black line in the bottom part: flux residuals after subtracting the 
quiescent spectrum from the active spectrum.
\label{fig:excessflux}}
\end{center}
\end{figure}

The flare phase is characterized by an increase in the core emission of
the chromospheric lines. Figures~\ref{fig:excesCaii} and \ref{fig:excessflux}
show the line cores of the {\ion{Ca}{II} K} line and the shortest-wavelength
component of the \catr\ IRT (8498\,\AA) at two different times, namely in the quiescent
phase and during the more active phase in the second half of the observation.
In the bottom parts of the figures we explicitly show
the flux residuals, demonstrating that the core emission is clearly stronger
during the later observation for both lines.

To study the evolution of the line cores we subtracted the quiescent reference
spectrum from each observed spectrum and thus obtain the residuals in the line
cores. Subsequently, we fit these residuals with a Gaussian profile
and determine the corresponding excess equivalent width $W_{\mathrm{ex}}$ for
all spectral lines under consideration. Since our spectra are flux
calibrated, we could directly convert measured excess equivalent width into the
flux observed at Earth by multiplying $W_{\mathrm{ex}}$ by the reference
continuum flux around the line.
Because of the continuum flux evolution in the flux calibrated spectra during
the observation run (see Fig.\,\ref{fig:SeeingAirmass}), we used an average 
value for the continuum reference flux based on 
all 244 spectra.

We carried out this procedure for all the selected lines and the
resulting light curves are depicted in 
Fig.\,\ref{Exc:evolution}. Qualitatively, all but one of the lines
show the same behavior as the \cahk\ line cores (cf.,
Fig.~\ref{fig:rhkIndex}), namely, an abrupt rise around the center of the 
exoplanetary transit followed by a slower decay. In the case of the \cahk, \catr\
IRT, \ha, and \hb\ lines also the secondary and tertiary event can be
identified.
Only the \hei\ line shows a somewhat different behavior with
hardly any excess emission during the rise phase and a turn into
absorption during the decay phase of the primary event; this behavior is
discussed in detail in Sect.~\ref{Sec:hed3}.

Based on the core emission light curves shown in Fig.\,\ref{Exc:evolution},  
we determine the start of the primary flare and
the time of the peak flux for each line individually (we followed the procedure
described in Sect.~\ref{Sec:rhk}). The results are indicated in
Fig.~\ref{Exc:evolution} and the 
values are summarized in Table~\ref{tab:FlareStart}.

The gradual decay phases of the primary event observed in the Balmer lines can
be described by an exponential decay function. In the \hg, \hde, and \he\ lines
no secondary and tertiary event can be identified so that a single exponential
is sufficient to model the decay phase. In the case of \ha\ and \hb, we modeled
the decay phase using a combination of
three exponential decay functions representing the primary, secondary, and
tertiary events.

In the case of the \catr\ lines, the decay of the primary event is not well
represented by an exponential. Therefore, we modeled the decay of the primary by
a polynomial and superimpose two exponentials representing the secondary and
tertiary events.
The resulting fits for all lines are
depicted in Fig.~\ref{Exc:evolution}. Based on our
modeling, we determine the e-folding times for all events and spectral lines
and the peak fluxes for the secondary and tertiary flares. The
resulting values are listed in Table~\ref{tab:FlareTime}.

\begin{table*}[t]
\begin{center}
  \caption{Energy budget of the flare.
  \label{tab:FlareFlux}}
  \begin{tabular}{l l c c l c} \hline\hline
\multirow{2}{*}{Line} & $E_{\mathrm{rise}}$  & $E_{\mathrm{SF}}$ &
$E_{\mathrm{TF}}$ &  $E_{\mathrm{Tot}}$  & $L_{\mathrm{PF}}$\\
 & [$10^{30}$\,erg] & [$10^{27}$\,erg] & [$10^{27}$\,erg] & [$10^{31}$\,erg] & [$10^{27}$\,erg s$^{-1}$]\\
\hline
\ha  &        $1.82$  & 4.74 & $3.14$ & $1.67$ & 3.11\\
\hb  &   $1.81$       & 3.60 &  2.74  &$1.56$ & 2.18\\
\hg &   $0.91$      & -- &  --  &0.45 & 1.15\\
\hde &   $0.26$      & -- & --  & 0.37 & 0.95\\
\he  &   $0.64$   & -- & -- &0.35 & 0.79\\
\hline
\catr\ K   & $1.33$&  4.21  &  0.89  &1.17 &  1.80   \\
\catr\ H   &$1.17$ & 2.81   &  0.55   &1.00 & 1.55     \\
\hline  
\catr\ IRT $\lambda 8498$  &  $0.79 $   & 2.10 &   0.31   & $0.55$ &  0.77  \\
\catr\ IRT $\lambda 8542$  &  $1.06$    & 0.55 &  1.16    & $0.82$ & 1.12\\
\catr\ IRT $\lambda 8662$  &  $0.71$   & 1.22 &  0.51      & $0.75$ & 1.04\\ 
\hline\\[-2ex]
Total energy & $10.54$ & -- & -- &$8.69$ &--\\
\hline
\end{tabular}
\tablefoot{$E_{\mathrm{rise}}$ - energy released during the impulsive phase;
$E_{\mathrm{SF}}$ - energy released by the secondary flare;
$E_{\mathrm{TF}}$ - energy released by the tertiary flare;
$E_{\mathrm{Tot}}$ - energy released during the total flare; $L_{\mathrm{PF}}$ - 
peak luminosity of the primary flare. $1\,\sigma$ error estimation is within 
$4\%$ for all the values.}
\end{center}
\end{table*}

\begin{table*}[t]
\begin{center}
  \caption{Time characteristics of a flare in studied lines. 
  \label{tab:FlareStart}}
  \begin{tabular}{l c c c c} \hline\hline
  \multirow{2}{*}{Line} & $S_{\mathrm{PF}}$ & $M_{\mathrm{PF}}$ & $M_{\mathrm{SF}}$ & $M_{\mathrm{TF}}$\\
  & [MJD-56109] & [MJD-56109] & [MJD-56109] & [MJD-56109]\\
\hline\\[-2.0ex]
   \ha & $0.2611 \pm 0.0006$ &  $ 0.2775 \pm 0.0004$  & 
$0.3024^{+0.0003}_{-0.0004}$ & $0.3288^{-0.0003}_{+0.0004}$\\[3pt]
   \hb & $0.2595  \pm 0.0010$ &  $0.2790 \pm 0.0006$  & 
$0.3022^{+0.0006}_{-0.0001}$ & $0.335^{+0.002}_{-0.004}$\\[3pt]
   \hg & $0.2601  \pm 0.0014$ &  $0.2795 \pm 0.0010$  & -- & --\\
   \hde & $0.2602  \pm 0.0010$ &  $0.2716 \pm 0.0010$ & -- & --\\
   \he &  $0.2560 \pm 0.0014$ &  $ 0.2787\pm 0.0009 $ & -- & --\\
\hline\\[-2.0ex]
   \catr\ K  &  $0.2604\pm 0.0006$   &      $0.2794 \pm 0.0005$    &  
$0.3086^{+0.0297}_{-0.0080}$ & $0.3217^{+0.0230}_{-0.0214}$ \\[3pt]
   \catr\ H  &  $0.2601 \pm  0.0007$ &      $0.2792 \pm 0.0006$    &   
$0.3066^{+0.0269}_{-0.0066}$ & $0.3232^{+0.0187}_{-0.0230}$   \\[3pt]
\hline\\[-2.0ex]
\catr\ IRT $\lambda 8498$    &  $0.2642 \pm  0.0008$ & $0.2838 \pm 0.0010$   & 
$0.3173^{+0.0268}_{-0.0172}$  & $0.3210^{+0.0249}_{-0.0204}$\\[3pt]
   \catr\ IRT $\lambda 8542$    &   $0.2601 \pm 0.0010$ & $0.2823 \pm 0.0009$ &  
$0.3046^{+0.0053}_{-0.0033}$ &  $0.3235^{+0.0053}_{-0.0034}$\\[3pt]
   \catr\ IRT $\lambda 8662$    &   $0.2624 \pm 0.0011$ & $0.2795 \pm 0.0012$ & 
$0.3040^{+0.0029}_{-0.0039}$ & $0.3255^{+0.0233}_{-0.0085}$ \\[3pt]   
\hline\\[-2.0ex]
\end{tabular}
\tablefoot{$S_{\mathrm{PF}}$ - start of the primary flare with $1\sigma$ confidence 
limits; $M_{\mathrm{PF}}$ - peak of the primary flare/the end of the impulsive 
phase; $M_{\mathrm{SF}}$ 
- peak of the secondary flare; $M_{\mathrm{TF}}$ - peak of the tertiary flare. 
Calculation of $1\sigma$ errors for $S_{\mathrm{PF}}$ and $M_{\mathrm{PF}}$ are 
based on \citet{Filliben1972}. Errors of $M_{\mathrm{SF}}$ and 
$M_{\mathrm{TF}}$ are determined by MCMC method.}
\end{center}
\end{table*}

\begin{table*}[t]
\begin{center}
  \caption{Flare duration and e-folding time.
  \label{tab:FlareTime}}
  \begin{tabular}{l c c c c c} \hline\hline
  \multirow{2}{*}{Line} & $\Delta t_{\mathrm{rise}}$ & $\Delta t_{\mathrm{F}}$ &  $\tau_{\mathrm{PF}}$  & $\tau_{\mathrm{SF}}$ & 
$\tau_{\mathrm{TF}}$ \\
   & [min] & [min] & [min] & [min] & [min]\\
  \hline\\[-2.0ex]
\ha  &23.62 & 144.16 & $119.1^{+3.7}_{-9.2} $& 
$10.3^{+1.8}_{-2.1}$ & 
$9.25^{+0.75}_{-1.15}$\\[3pt]
\hb  &   28.08  & 146.47 & $90.8^{+5.1}_{-5.5}$ & 
$9.36^{+0.35}_{-0.29}$ & 
$7.56^{+0.44}_{-0.43}$\\[3pt]
  \hg  &  27.94 & 145.60 &$56.5^{+5.4}_{-5.0}$ & -- &--\\[3pt]
   \hde & 16.42  & 145.44 &$70.2^{+8.5}_{-7.2}$ &-- &--\\[3pt]

  \he  & 32.69 & 151.51 &$72.8^{+9.5}_{-9.8}$ & --&--\\[3pt]
  \hline\\[-2.0ex]
  \catr\ K  & 27.36 & 145.17 & 121.43 & $9.7^{+1.3}_{-1.3}$& 
$8.2^{+1.4}_{-1.2}$\\[3pt] 
  \catr\ H  & 27.50 & 145.60 & 121.53 & $11.0^{+1.7}_{-1.9}$ & 
$8.0^{+1.4}_{-1.0}$\\ [3pt]
  \hline\\[-2.0ex]
  \catr\ IRT $\lambda 8498$  & 28.22 & 139.70 & 119.88 
& $10.9^{+1.8}_{-1.9}$ & $ 7.6^{+1.8}_{-1.6}$\\ [3pt]
  \catr\ IRT $\lambda 8542$  &  31.97  & 145.60 & 140.22 
&$10.8^{+1.9}_{-1.8}$ & $8.7^{+1.3}_{-1.8}$\\ [3pt]
  \catr\ IRT $\lambda 8662$  & 24.62 & 142.29 & 137.72 
&$11.1^{+1.9}_{-1.7}$ & $7.9^{+1.7}_{-1.9}$\\[3pt]
  \hline
  \end{tabular}
\tablefoot{$\Delta t_{\mathrm{rise}}$ - duration of an impulsive phase; $\Delta t_{\mathrm{F}}$ - 
duration of a whole flare; $\tau_{\mathrm{PF}}$ - e-folding time of the primary
flare; $\tau_{\mathrm{SF}}$ - e-folding time of the secondary flare;
$\tau_{\mathrm{TF}}$ - e-folding time of the tertiary flare. MCMC method was 
used for the error determination.}
\end{center}
\end{table*}

\subsection{\hei}
\label{Sec:hed3}

The excess equivalent width $W_{\mathrm{ex}}$ in the core of the \hei\ line
shows a different evolution during the 
observation than the other studied lines (see Fig.~\ref{fig:Hed3ew},
bottom panel). While there may be a small amount of excess emission during the
flare rise phase observed in other lines ($\sim 0.2\times
10^{-14}$~erg\,cm$^{-2}\,s^{-1}$), there is no clear flare signature, and the
line goes into absorption compared to the pre-flare level during the
decay phase of the primary event.

Based on this finding, we checked the behavior of the \hei\
line as a whole by computing its equivalent width, whose evolution 
is shown in the top panel of Fig.~\ref{fig:Hed3ew}. It confirms the diverse
behavior of the line. During quiescence a \hei\
EW of $\sim 13.5$~m\AA\ is measured, corresponding to a surface flux, $F_{D3}$,
of $51\,000$~\mbox{erg\,cm$^{-2}$\, s$^{-1}$}. These values are likely slightly
overestimated because the line is blended \citep[e.g.,][]{Saar1997}.
The EW remains essentially
constant until approximately 35\,min after the mid-transit time and rises afterward.
About 70\,min after the mid-transit time the equivalent
width seems to stop rising and remains constant until the end of our
measurements.
The increase in equivalent width is about 1\,m\AA. For reference, we reproduced
the evolution of the excess equivalent width, $W_{\mathrm{ex}}$, in the core of
the \hei\ line in Fig.~\ref{fig:Hed3ew}. Both plots show
that the absorption in the \hei\ line increases compared to the quiescent level
in contrast to the other studied chromospheric lines.

To rule out instrumental effects, we
checked neighboring spectral regions and studied the 
equivalent width of three selected photospheric lines and compare the
evolution of continuum regions without strong spectral lines.
The resulting light curves are plotted in Fig.~\ref{fig:Hed3photlines}.
They are essentially constant during
the observation. None of
them shows behavior similar to the \hei\ line. Therefore, we argue that the
observed behavior in the \hei\ line is not instrumental.

\subsection{Chromospheric flare energy budget}\label{sec:EnergyBudget}

Based on our excess flux measurements, we calculated the energy released during
the flare in the individual lines. We converted the measured flux at Earth,
$F_{\oplus}$, into luminosity, $L$, using the relation
\begin{equation}
L=4\pi d^2 \cdot F_{\oplus}
\end{equation}
where $d$ is distance to \hd\ (see Table~\ref{tab:Properties}).

For the primary event, we determined the energy released during the rise 
phase by integrating the chromospheric excess emission during that phase.
Similarly, the energy released during the decay phase was obtained by
integrating the measured flux in that phase. 
As is clear from Figs.\,\ref{fig:rhkIndex} and \ref{Exc:evolution},
the primary event is not completely covered by our observations.
To obtain an estimate of the missing fraction, 
we extrapolated the evolution of the flare and estimate that about $10$\% of the
total energy released in the \cahk\ lines remain unobserved.
As the estimate of the energy released by the primary event is, however,
further complicated by the uncertainty in the definitions of the quiescent level
estimation, which is about 2\%, the
definition of the start of the flare, and the uncertainty in
the flux calibration, which is on the order of 20\% 
(this value is based on the comparison of our flux 
calibrated spectra with the predicted Kurucz model 
atmosphere as discussed in Sect.~\ref{Sec:obsandred} of this 
paper and in Sect.~5.2 in \citet{Czesla2015}),
we argue that the missing energy fraction does not significantly influence
the resulting overall energy budget and can, therefore, be neglected in the
following analysis.

Adding the energy released during the rise and decay phases, we calculated
the total energy released during the observed flare event in the individual
lines. In the case of the secondary and tertiary flares, the peak fluxes, decay times,
and total energies can be directly derived from our modeling of the excess flux
by an exponential. The  statistical ``1 $\sigma$'' error estimate for the energy
measurements (for all the lines) is on the order of a few percent from the
determined values.
Our results are summarized in 
Tables~\ref{tab:FlareFlux} and~\ref{tab:FlareTime}.

In the case of the \cahk\ lines, an alternative estimate of the peak flux and
released energy can be obtained based on $R'_{\mathrm{HK}}$.
Combining the effective temperature and stellar radius given in
Table~\ref{tab:Properties} and the quiescent level of $-4.525$ for
$R'_{\mathrm{HK}}$ estimated in Sect.~\ref{Sec:rhk}, we estimated a quiescent
chromospheric \cahk\ line luminosity of $3.85\times 10^{28}$\,\ergs\ for \hd. 
The peak flux for the primary event indicated in
Fig.\,\ref{fig:rhkIndex} yields a value of $2.06\times10^{27}$~\ergs\ for the
corresponding peak luminosity. By integrating the energy released
by the primary event in excess of the quiescent level, we deduced a value of
$1.26\times 10^{31}$~erg for the total energy
released in the \cahk\ emission line cores by the primary event. This value is 
smaller by about $70$\% than the overall energy released in 
\cahk\ lines $E_{\mathrm{Tot}}=2.17\times 10^{31}$\,erg determined with the 
method described in Sect.~\ref{Sec:exces_cores}. The two values are nevertheless 
on the same order and this discrepancy can be attributed to the imperfection in 
the flux calibration and the derivation of the $R'_{\mathrm{HK}}$ index.

\section{Discussion}\label{Sec:discussion}

The temporal evolution in the cores of the chromospheric lines displayed in
Fig.\,\ref{Exc:evolution} shows a stellar flare, beginning shortly
after mid-transit time.
While the line evolution is qualitatively similar
for all the lines except for the \hei\ line, a more detailed inspection shows
quantitative differences between individual lines.

\subsection{Evolution of the Balmer and \catr\ lines}

Both the Balmer and \catr\ lines show a 
steep rise phase, followed by a much slower decay phase.
The measured durations of the rise phase in all the studied lines are
in the range of 20-30\,min with a median of $28$\,min.
The peak and subsequent decay clearly differs between the Balmer lines and the
\catr\ lines. While all Balmer lines show a sharp flare peak, immediately
followed by a decay, which is well described by an exponential function,
all \catr\ lines lack the sharp flare peak, rather they show
a ``plateau'' after the rise phase. Therefore, an accurate definition of
the instant of peak flux is difficult in the case of the \catr\ lines.

While the \ha, \hb, \hg, and \he\ lines all peak within about three minutes, the
estimated peak of the \hde\ line occurs approximately $10$~min before that
(see Table~\ref{tab:FlareStart}).
A close visual inspection of the \hd\ light curve shows, however, that
the adopted peak position mainly depends on two data points and the physical
peak could also plausibly occur later when the remaining Balmer light curves
peak. This appears even more plausible considering that we determined the by far
shortest rise phase of $16.4$\,min for the \hde\ line.
We thus consider the evidence insufficient to claim a significant delay
in the flare peak observed in the Balmer lines.

In the \catr\ lines, we identified the instant of peak flare flux with the
beginning of the plateau phase.
According to our definitions, there is a tendency for the
\catr\ lines to peak a few minutes after the Balmer lines. For example,
the \ha\ line peaks approximately 2.5~minutes before both \cahk\ lines.
In an analysis of several flares on AD~Leo, \citet{CrespoChacon2006}
find that the flare peaks in the \cahk\ lines were delayed with respect to
that of the Balmer lines by up to $5\pm 3$~min; in particular, this refers to
the delay of the \catr~K line with respect to \hb. If anything, our observation
indicates a similar trend.
However, the morphology of the \catr\ light curves prevents an accurate
identification of the start of the plateau phase. So while the
\catr\ lines definitely show a different temporal evolution than the Balmer lines, a
delay in the instant of peak flux cannot be uniquely detected.

The delay in the range of 5-20\,min of the flare peak in 
\cahk\ lines 
with respect to Balmer lines is frequently observed in flares on M-type stars, 
\citep[e.g.][]{GarciaAlvarez2002,Fuhrmeister2008,Hawley1991}, on the other hand, 
there are also observations of flares, where Balmer lines peak together 
with the \catr\ lines \citep[e.g.][solar flare]{Kowalski2015}. The evolution of the
optical chromospheric lines during a flare show an apparently diverse behavior 
and is a subject of ongoing research. There have been several attempts to 
explain the mechanism(s) responsible for this observational feature. \citet{Hawley1992} 
present several models to investigate whether X-ray heating from the 
corona can explain the chromospheric emission observed in the optical region. 
\citet{Houdebine2003} model the behavior of the Balmer and \catr\ lines, 
applying a natural radiative cooling of the flare plasma mechanism. In this 
scenario the rise and decay times of the flare depend on the density of the 
plasma, which predicts a faster evolution of the flare in the high-temperature lines - 
an effect observed during several stellar flares and confirmed by 
us as well (see the discussion in the following paragraph).

Since the end of the flare is beyond the end of our observing run, we 
can only determine a
lower limit of $\sim 145$\,min for the duration of the entire event. However,
the major fraction of the flare is certainly covered so that the
decay timescales in the individual lines can be accurately estimated. 
From our exponential fits, we find that the decay times 
decrease from \ha, over
\hb, to \hg. In our sample, the latter shows the shortest decay time of
$56.5$\,min while the \hde\ and \he\ lines show longer decay time of about
$70$\,min. We argue that the unusually short decay time of \hg\ is likely
spurious, potentially attributable to the unusual behavior of its light curve
during the last quarter of the observation, where it seems to settle
near the quiescence level. The e-folding times of the \catr\ light curves are 
comparable
to that of the \ha\ line and longer than that of the other Balmer lines. It is,
however, not well described by an exponential after the rise phase, where it
shows a plateau and evolves more gradually. The \cahk\ line fluxes reach
smaller peak fluxes than the \ha\ line, but decay more slowly so that \ha\ and
\catr~K reach about the same level of excess emission at the end of our
observation.

In their analysis of a large flare on the active M2 dwarf AD~Leo,
\citet{Hawley1991} also find a sharper peak
and faster decay in the \hg\ and \hde\ lines compared to the \catr~K
light curve, which shows a ``rounder'' evolution after the impulsive phase (see
their Fig.~7). In the flare studied by \citet{Hawley1991}, the peak fluxes
reached by the \hg\ and \hde\ lines are about three times larger than the peak
flux reached by \catr~K. In our case, the peak flux of the strongest
Balmer line (\ha), is only about $40$\% higher than that observed in \catr~K,
indicating a somewhat different distribution of energy among the chromospheric
lines.

During the decay phase of the primary flare,
we detect a smaller secondary and a tertiary flare showing decay
times on the order of $10$\,min, which contribute a few percent of the radiated
energy in the \catr\ and Balmer lines, where they could be detected. 
Based on our data, it remains impossible to prove a physical relation between
these smaller fares and the primary event. Yet, no comparable flare event was
observed in the first half of the observation, while we detect three flares in
the second half. The existence of these secondary events is
in agreement with the prediction of the standard flare model
\citep[e.g.,][]{Kopp1976}. Major flares can trigger secondary flares
in neighboring magnetic loop systems and
also in remote active regions, which may be magnetically connected
\citep[so called sympathetic flares, e.g.,][]{Moon2002, Wang2007}.

\subsection{The behavior of \hei}

The \hei\ line(s) are formed by transitions between triplet states more than
$20$~eV above the ground state, which are not easily populated at photospheric and
chromospheric temperatures.
Therefore, the \hei\ line is weak or absent in the spectra of inactive stars,
and its presence in the spectra of cool stars points to the existence of
non-radiative heating processes \citep{Saar1997}.
Two main mechanisms are discussed to populate the
triplet states, viz., photoionization by coronal X-ray and extreme ultraviolet
radiation followed by radiative recombination and
collisional excitation in the hot upper chromosphere or transition region
\citep[$\sim 20\,000$~K, e.g.,][]{Athay1960, Zirin1975, Milkey1973}; see
also \citet{GarciaLopez1993} for a discussion, however, the simulations by
\citet{Lanzafame1995} indicate a minor role of coronal irradiation in the
formation of the He triplet lines.

Indeed, higher activity levels can lead to stronger absorption in the \hei\
lines by more densely populating the helium triplet states.
\citet{Saar1997} argue that the behavior of the line is similar
to the behavior of the \ha\ line in cooler M~dwarfs. As a consequence of rising
chromospheric activity levels in very cool stars, the $n=2$ energy level
becomes more populated and the equivalent width of \ha\ absorption increases.
At the point where the level populations become collisionally controlled,
the source function is determined by the temperature and as a result the line
will turn into emission \citep[e.g.,][and references therein]{Robinson1990,
Cram1987}.

According to \citet{Saar1997}, the behavior of the \hei\ line in active G and K dwarfs
correlates with other activity indicators, such that, for instance, higher
levels of \cahk\ line emission correspond to stronger \hei\ absorption. 
The same trend is observed for \hd\ by \citet{Boisse2009}, who study the
evolution of \cahk, \ha, and \hei\ indices during quiescence.
We find that our results are consistent with the relations between helium flux,
$F_{D3}$, and stellar color, rotation period, and \cahk\ flux determined
in the sample analysis by \citet{Saar1997}\footnote{From the conversion between EW and helium
flux it appears to us that the fluxes given by \citet{Saar1997} refer to
astrophysical fluxes, which differ from physical fluxes by a factor of
$\pi^{-1}$.} (see their Figs.~3, 4, and 6). Our value 
of 
$F_{D3}=16,000$\,erg\,cm$^{-2}$\,s$^{-1}$ agrees well with the range of 
$F_{D3}$ values determined by \citet{Saar1997} for K1-2V type stars (cf., their 
Tab.~1).

On the Sun, the \hei\ line is seen in emission only in more intense flares and
in absorption in eruptive prominences and weaker flares
\citep[][Sect.~11.3]{Zirin1988}.
Early solar studies have shown that the peak instants of flares observed in \hei\
can occur either at the same time as that of the Balmer lines or be delayed by
up to 20~minutes \citep[e.g.][and references therein]{Bray1964}.
\citet{GarciaAlvarez2005} study the evolution of the \hei\ line during a 
medium, GOES C7.5 class solar flare and find that
the \hei\ line remains in absorption for the whole event, and its strength is
correlated with that of the \ha\ and \catr~K lines. Similarly,
\citet{JohnsKrull1997} report on a GOES M7.7 class solar flare also finding
the \hei\ line to be in absorption throughout the entire event and absent during 
quiescence. However, the authors detect a constant level of absorption in \hei\ 
for about $40$~min after the flare peak observed, for example, in H$\alpha$ and soft 
X-rays. 
Only after that the \hei\ line EW starts to drop and follows the decay of the other
chromospheric activity indicators.

During flares on other stars the \hei\ line has been observed in emission. For
instance, \citet{CrespoChacon2006} detect the line studying flares on the
active M-type star AD~Leo and \citet{Montes1999} report \hei\ emission during
a flare on the K2 dwarf LQ~Hya. Our observations indicate a potential small
increase in the \hei\ flux during the rise phase of the flare, which would be
consistent with observations of strong flares on the Sun,
thereafter a phase of increasing absorption follows.
We speculate that the latter may, indeed, be caused by the flare, which may
lead to a stronger population of
the excited \ion{He}{I} levels, by providing additional energy to the
stellar atmosphere both in the form of local heating and more globally through
the increased level of extreme-ultraviolet and soft X-ray emission illuminating the
flaring hemisphere. This hypothesis would gain strong support from a return of
the \hei\ absorption levels to the pre-flare level after the event.
Unfortunately, the event has not been fully covered and during our observation no
reversal of the trend in the \hei\ could be observed. Therefore, any conclusion
on the nature of the observed behavior of the \hei\ line remains speculative.

\begin{figure*}
\includegraphics[width=\textwidth]{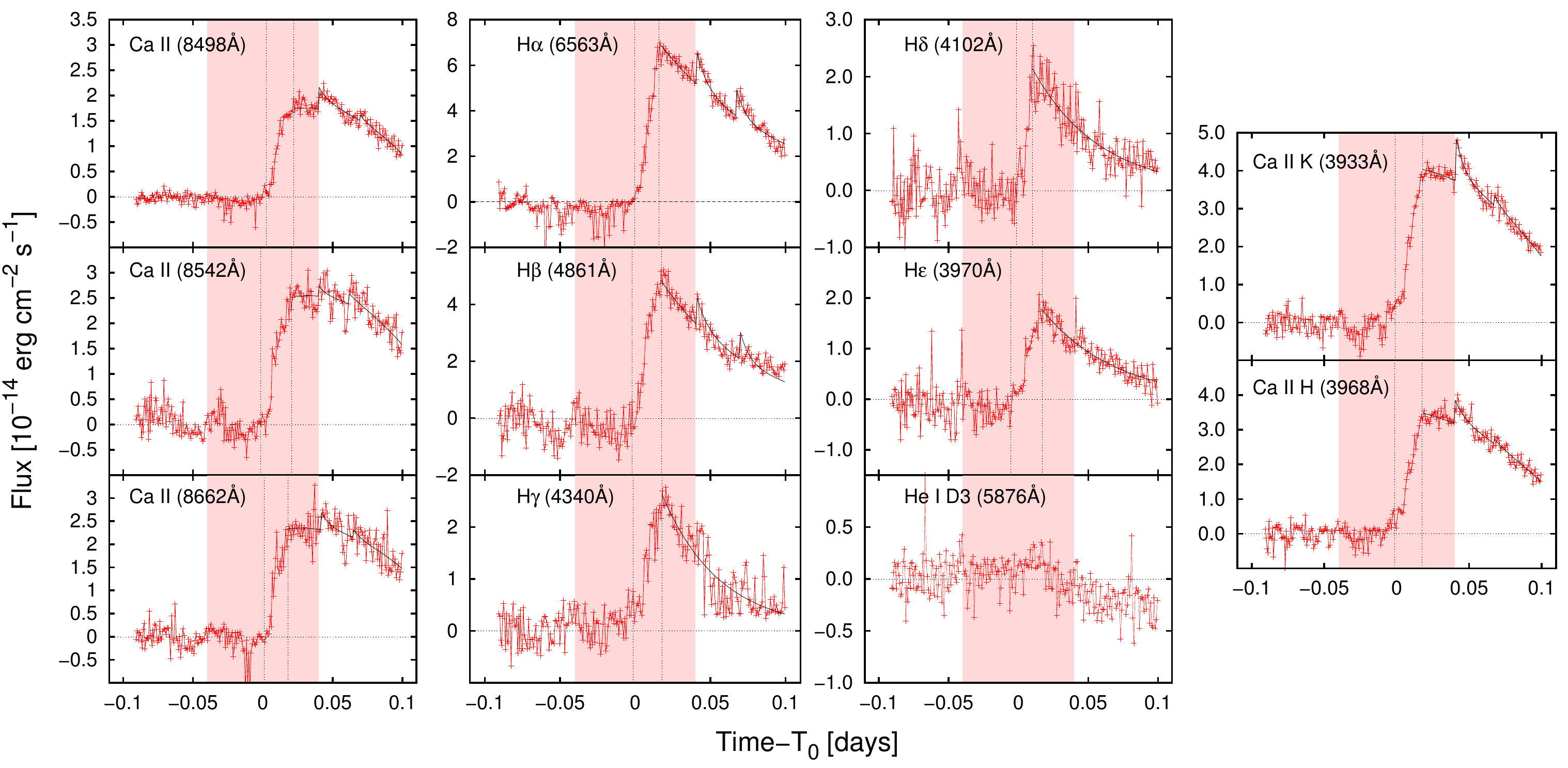}
\caption{Evolution of a flare in selected chromospheric optical lines. The $y$ axis is the flux measured at Earth, $x$ axis is time 
in days. The center of the 
$x$ axis (time=0\,days) corresponds to the middle 
of the transit ($T_{0}$\,$[\mathrm{HJD,UTC}] = 2456109.26164$). Red crosses are 
the measured flux value. Full black lines represent a best fit of the flare 
decay phase. Dashed black vertical lines 
mark a beginning and 
peak of the flare for each line individually. 
Horizontal dashed black lines indicate a quiescent zero 
level. The shaded pink area labels the 
duration of an exoplanetary transit. The 1\,$\sigma$ 
error for the 
displayed data points is within the 3-6\% range.}
\label{Exc:evolution}
\end{figure*}

\subsection{Flare energy budget}

The total energy released by the flare 
within all the studied chromospheric lines is $E_{\textrm{tot}} = 8.7\times
10^{31}$\,erg (see Table~\ref{tab:FlareFlux}). 
The largest contribution is due to the Balmer lines, followed by the
\cahk\ lines. During a large flare on AD~Leo studied by \citet{Hawley1991} only
between $10$\% and $40$\% of the energy radiated in the $1200-8000$\,\AA\ band,
is emitted in the form of emission lines. Therefore, we estimate that the true
energy budget is probably larger by a factor of between two and ten.

A number of optical spectroscopic observations of flares on K-type stars
are reported in the literature 
\citep[e.g.,][]{Montes1999, LopezSantiago2003, Montes2005}.
\citet{Montes1999} study a long-duration ($\> 2$~h) flare observed in the
optical and ultraviolet regime on the young, single K2 dwarf LQ~Hya. The total
energy released in the optical chromospheric lines studied by \citeauthor{Montes1999}
(\ha, \hb, \hei, \mg\ triplet, \nad, \fe\ and \hel\ $\lambda 6678$) is $\geq
5.7\times 10^{33}$\,erg, that is, two orders of magnitude higher than the value
determined for the flare on \hd. Accordingly, the peak \ha\ flux
is also two orders of magnitude higher.
Although ``our'' flare on \hd\ is comparable to or larger than the strongest 
solar flares, which release
up to $10^{32}$\,erg \citep{Priest2014}, it certainly remains moderate in the context
of flares observed in young K dwarfs and dMe stars, which
release up to about $10^{34}$\,erg \citep{Hawley1991}.

For a number of flares on dMe stars, \citet{Hawley1991} find that the energy
released in \catr~K line is about half of that released in the \hg\ line. This is
not the case in the flare observed on \hd. In fact, we find approximately the
inverse ratio with about $2.5$ times the energy of the \hg\ line released in the
\catr~K line. At this point, we cannot determine whether this difference is a
particularity of this specific flare or more generally valid for flares on early
K-type stars such as \hd.

\subsection{Extrapolating into the X-ray regime}
\hd\ is a known X-ray source with a quiescent X-ray luminosity of about
$1.1\times 10^{28}$~erg\,s$^{-1}$ \citep{Poppenhaeger2013}, and several X-ray
flares have been observed on the star \citep[e.g.,][]{Pillitteri2014}. 
Although no simultaneous X-ray data are available for the flare studied
here, we attempt a rough estimate of the level of the soft
X-ray flux during the flare using the correlations between the \cahk\ flux
and the soft X-ray flux given by \citet{Schrijver1983}.

Combining our \cahk\ measurements with Eq.~6 given by
\citet{Schrijver1983} ($F_{x}=3.4\cdot 10^{5}\Delta F^{1.67}_{H+K}$), we 
estimate a value of
\mbox{$3.2 \times 10^{5}$\,erg cm$^{-2}$ s$^{-1}$} for the quiescent X-ray 
surface flux
of \hd, which compares well with the nearly identical observed value. While we
therefore find the relation to also be valid in the case of \hd,
we consider the excellent numerical correspondence a coincidence.

During the flare, the emission in the \cahk\ lines increases by about $6$\%.
Assuming that the \citeauthor{Schrijver1983} relation, $F_X \sim \Delta
F_{H+K}^{1.67}$, can also be applied during the flare, we estimated a peak soft
X-ray luminosity of $10^{27}$~erg\,s$^{-1}$ for the flare. Further assuming
the same e-folding time of $120$~min as in the \cahk\ lines also in X-rays,
we estimated that a total energy of $7\times 10^{30}$~erg
was released by the flare in soft X-rays.
Alternatively, we could apply the relation, $L_X = 31.6 \times L(\hg)$, derived
from flare observations on nearby dMe stars \citep{Butler1988} and arrive
at an estimate of $14\times 10^{30}$\,erg for the total energy radiated at soft
X-rays, which agrees with the former estimate to within a factor of two.
According to these estimates, the X-ray
flares reported on by \citet{Pillitteri2014} are about an order of magnitude
stronger with mean flare luminosities of several times
$10^{27}$~erg\,s$^{-1}$ and total energies of several times $10^{31}$~erg
released in X-rays.

As the planet and the flare were located on the same stellar hemisphere,
the planetary atmosphere was clearly exposed to the full flare radiation.
However, we consider a transient rise in X-ray emission by $\sim 10$\% too 
small to significantly
affect the hydrodynamic escape of the planet's atmosphere, which can only 
use about $1$\% of the irradiated energy to drive atmospheric 
escape \citep{Salz2016}. Yet,
significantly stronger flares could, indeed, lead to a significant change in
the atmospheric structure \citep[][]{Lecavelier2012} and the impact of
many smaller flares may also accumulate.

\subsection{Star-planet interaction}

As discussed in Sect.~\ref{intro}, the activity 
and 
proximity of \hd\ and its planet make this system an ideal target for 
studies of star-planet interaction (SPI). SPI was theoretically predicted 
by 
\citet{Cuntz2000} 
and was later claimed by, for example, \citet{Shkolnik2005,Shkolnik2008} as 
a 
modulation of selected lines in the spectra of stars with close-in planets 
phased with the planetary orbit. SPI is still a controversial subject and 
a topic of much ongoing research. 

\citet{Fares2010} use spectropolarimetric measurements of this 
system covering two stellar rotation cycles and find that the
residuals of the \ha\ and \cahk\ lines with respect to an average 
line profile are periodic and modulated with the rotation period of 
the star, rather the the orbital period of the planet; hence \citet{Fares2010} 
do not find any evidence for SPI.
\citet{Cohen2011} perform time-dependent MHD modeling of SPI in the \hd\ system and 
show that in their simulations reconnection events can occur during a specific 
orbital phase. Such events could then be responsible for the mass loss from the planetary 
magnetosphere, causing a hotspot on the surface of the star observed as 
modulation in \cahk\ lines.
As described in Sect.~\ref{intro}, \citet{Pillitteri2010, Pillitteri2011, 
Pillitteri2014} observe flares occurring in \hd\ in quite a tight orbital phase 
interval, possibly caused by the orbiting exoplanet. Their findings are 
supported by the models of \citet{Lanza2012}, who predict a modulation of 
coronal flaring activity with the orbital phase of the planet.

In our case, we observe the flare starting in the middle of the primary 
transit. To our knowledge, so far there has been no observation of increased 
stellar activity during a primary transit. There are more than 20 observations 
of \hd\ during a primary transit 
\citep[e.g.][]{Redfield2008,Wyttenbach2015,Cauley2017}. Although flares were not specifically studied by these authors, 
it appears to be very plausible that an event of a similar magnitude to that observed by us would
have been noticed by the authors. Thus the hypothesis that this may be
a recurrent event seems unlikely. To prove this correlation, we would need more observation of a flare during a 
transit. Nevertheless, with respect to the simulations of \citet{Cohen2011}, it could 
have happened that the planet around \hd\ has encountered a locally enhanced 
magnetic field of a host star and induced a flare. So clearly more observations of
activity related features are required to arrive at firm conclusions with respect to to the
presence of SPI effects in \hd .

\begin{figure}
\begin{center}
\includegraphics[width=0.5\textwidth]{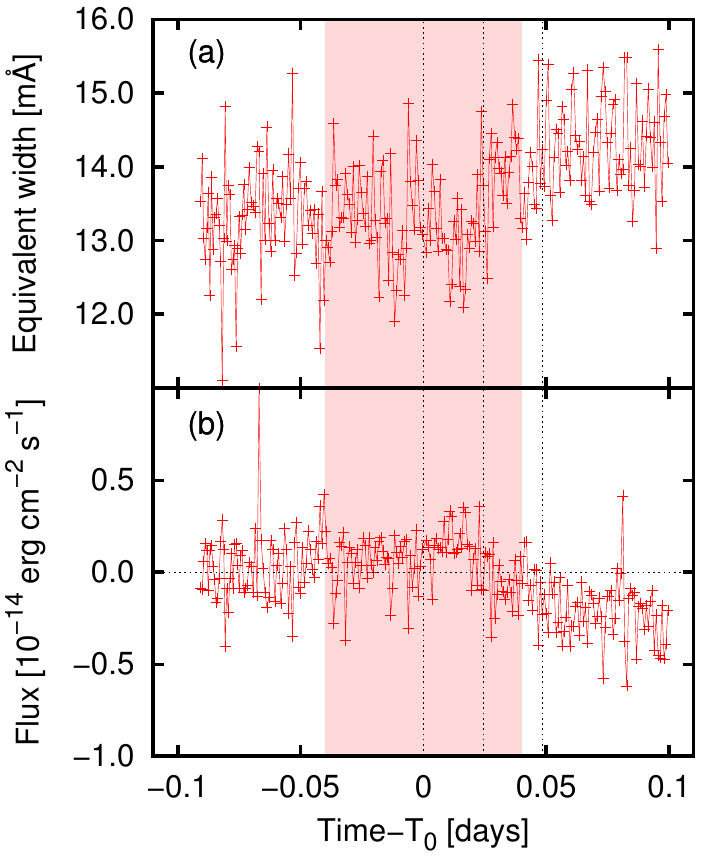}
\caption{Panel (a): Evolution of the \hei\ line equivalent width. Panel (b):
Evolution of the excess equivalent width, $W_{\textrm{ex}}$, in the core of the \hei\ line
converted into the flux measured at Earth (see also
Fig.~\ref{Exc:evolution}).
Shaded pink area indicates the duration of an exoplanetary 
transit. Meaning of dashed lines 
from left to right: Middle of the transit $T_0$\,$[\mathrm{HJD,UTC}] = 2456109.26164$; 
Start of the 
rise 
phase; End of the rise phase.
The 1\,$\sigma$ error for the displayed 
data points is within the 4-6\% range.
\label{fig:Hed3ew}}
\end{center}
\end{figure}

\begin{figure}
\begin{center}
\includegraphics[width=0.5\textwidth]{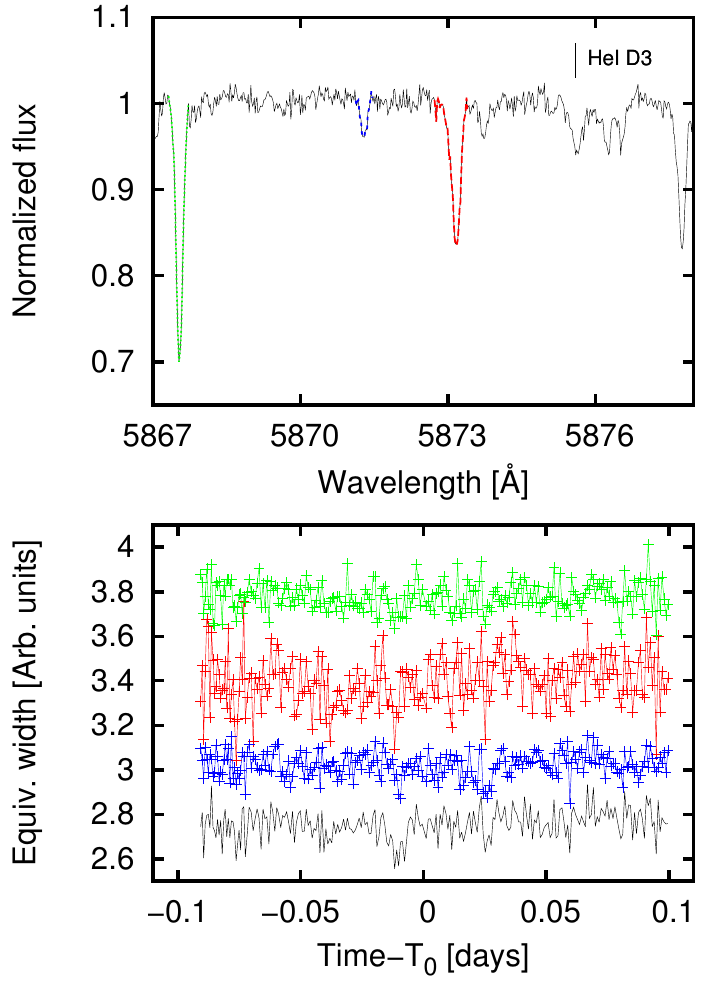}
\caption{{\it Top:} Spectral region around \hei\ line. Neighboring
photospheric lines used for the behavior check are marked with different colors.
{\it Bottom:} Evolution of the equivalent width of the selected photospheric
lines (marked with the same colors) and of a continuum in the vicinity of the
\hei\ line (black line). The curves were scaled for better comparison.
The center of the $x$ axis (time=0\,days) corresponds to 
the middle of the 
transit $T_{0}$\,$[\mathrm{HJD,UTC}] = 2456109.26164$.
\label{fig:Hed3photlines}}
\end{center}
\end{figure}

\section{Summary and conclusion}
\label{Sec:conclusions}
Using spectra obtained with the UVES spectrograph
we have presented a temporally resolved spectral analysis of the evolution in the
chromospheric lines \ha, \hb, \hg, \hde, \he, \hei, \cahk, and \catr\ IR triplet 
during a flare observed on the planet host star \hd; the flare
started shortly after mid-transit at $\sim$ MJD~56109.26, 
and finished 
after our observation.
To study the \cahk\ lines, we first used the chromospheric $\log R'_{\mathrm{HK}}$ activity index 
(Fig.~\ref{fig:rhkIndex}), subsequently,
we studied the excess emission in the cores of all the Balmer lines, the \cahk,
and \catr~IRT lines, and the \hei\ line relying on the flux calibration of our
spectra (Fig.~\ref{Exc:evolution}). The energetics and characteristics of the
flare are summarized in Tables~\ref{tab:FlareRhk}, \ref{tab:FlareFlux},
\ref{tab:FlareStart}, and \ref{tab:FlareTime}. 

The Balmer and \catr\ lines show a typical flare signature with a fast
impulsive phase with a duration of $\sim 28$\,min and a slower decay phase, which lasts for at 
least two hours. Nonetheless, there is a clear difference in the temporal
evolution of the Balmer and \catr\ lines with the former showing a distinct peak
followed by a quasi-exponential decay and the latter a more gradual evolution
after the impulsive phase. Qualitatively, this evolution is consistent with
previous flare observations in dMe stars.

The evolution of the \hei\ line follows a different temporal pattern than the
other chromospheric lines. It shows a potential small emission episode during
the impulsive phase of the flare and afterwards, with a
substantial delay, turns into absorption compared
to the pre-flare level. While this behavior may be related to the
flare, a unique association is difficult to establish based on the available data.

According to our analysis,
the total energy released in the studied optical chromospheric lines (except for
the \hei\ line) is $\sim 8.7\times 10^{31}$\,erg, which
puts the observed flare between the most energetic solar flares
and moderate stellar flares observed on K-type and M-type stars. 
We observed a smaller secondary and a tertiary flare occurring during the decay
phase of the primary event. Although it cannot be proved, they may well be
triggered by the primary flare.
As our observations remain so far the only in-transit flare observations in \hd,
although many observations during a transit have 
already been carried out
\citep[e.g.,][]{Redfield2008, Wyttenbach2015, Triaud2009}, we consider a
star-planet-interaction scenario as proposed by \citet{Pillitteri2014} to be unlikely
in this case and we rather consider the flare timing a chance coincidence.

The cores of selected optical spectral lines are used 
for the exoplanet atmosphere detection \citep[see e.g.][and 
references therein]{Charbonneau2002,Redfield2008}. Our 
analysis shows that the activity of an exoplanet host 
star can affect cores of a wide range of optical lines and 
must therefore be taken into account when studying 
exoplanetary atmospheres, as demonstrated 
by \citet{Khalafinejad2017}.

\begin{acknowledgements}
The authors would like to thank Prof.\,Petr Heinzel for 
helpful comments on the manuscript. This 
work has been supported by the DFG grants CZ 222/1-1 and SCHM 1032/37-1. TK and SK acknowledge 
support from the RTG 1351 ''Extrasolar planets and their host stars''. 
\end{acknowledgements}

\bibliographystyle{aa}
\bibliography{doc.bib}

\end{document}